\documentclass{article}

\usepackage{arxiv}

\usepackage[utf8]{inputenc} 
\usepackage[T1]{fontenc}    
\usepackage[hidelinks]{hyperref}       
\usepackage{url}            
\usepackage{booktabs}       
\usepackage{multirow}
\usepackage{amsfonts}       
\usepackage{nicefrac}       
\usepackage{microtype}      
\usepackage{lipsum}		
\usepackage{graphicx}
\usepackage{doi}
\usepackage{dirtytalk}
\usepackage{subcaption}
\usepackage{amsmath}
\usepackage{mathtools}
\usepackage{physics}
\usepackage{euscript}
\graphicspath{ {./images/} }

\title{FluxGAT: Integrating Flux Sampling with Graph Neural Networks for Unbiased Gene Essentiality Classification}

\author{\href{https://orcid.org/0009-0000-0336-9944}{\includegraphics[scale=0.06]{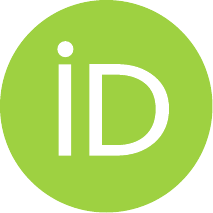}\hspace{1mm}Kieren Sharma}\thanks{Corresponding author. This work is currently under review. For questions or further information, please contact Kieren Sharma.}\\
	School of Engineering Mathematics and Technology\\
	University of Bristol\\
	Bristol, UK\\
	\texttt{\href{mailto:kieren.sharma@bristol.ac.uk}{kieren.sharma@bristol.ac.uk}}\\
 	\And
	\href{https://orcid.org/0000-0002-7553-6358}{\includegraphics[scale=0.06]{orcid.pdf}\hspace{1mm}Lucia Marucci}\\
	School of Engineering Mathematics and Technology\\
	University of Bristol\\
	Bristol, UK\\
	\texttt{\href{mailto:lucia.marucci@bristol.ac.uk}{lucia.marucci@bristol.ac.uk}} \\
	\And
	\href{https://orcid.org/0000-0002-1291-2918}{\includegraphics[scale=0.06]{orcid.pdf}\hspace{1mm}Zahraa S. Abdallah}\\
	School of Engineering Mathematics and Technology\\
	University of Bristol\\
	Bristol, UK\\
	\texttt{\href{mailto:zahraa.abdallah@bristol.ac.uk}{zahraa.abdallah@bristol.ac.uk}} \\
}

\begin{document}
\maketitle
\begin{abstract}
Gene essentiality, the necessity of a specific gene for the survival of an organism, is crucial to our understanding of cellular processes and identifying drug targets. Experimental determination of gene essentiality requires large growth screens that are time-consuming and expensive, motivating the development of \textit{in-silico} approaches. Existing methods predominantly utilise flux balance analysis (FBA), a constraint-based optimisation algorithm; however, they are fundamentally limited by the necessity of a predefined cellular objective function. This requirement introduces an element of observer bias, as the objective function often reflects the researcher’s assumptions rather than the cell’s biological goals. Here, we present FluxGAT, a graph neural network (GNN) model capable of predicting gene essentiality directly from graphical representations of flux sampling data. Flux sampling removes the need for objective functions, thereby eliminating observer bias. FluxGAT leverages the unique strengths of GNNs in learning representations of complex relationships within metabolic reaction networks. The success of our approach in predicting experimentally determined gene essentiality, with almost double the sensitivity of FBA, explores the possibility of predicting cellular phenotypes in cases when objectives are less understood. Thus, we demonstrate a method for more general gene essentiality predictions across a broader spectrum of biological systems and environments.
\end{abstract}

\section{Introduction}
\begin{figure*}[t!]
    \centering
    \includegraphics[width=\textwidth]{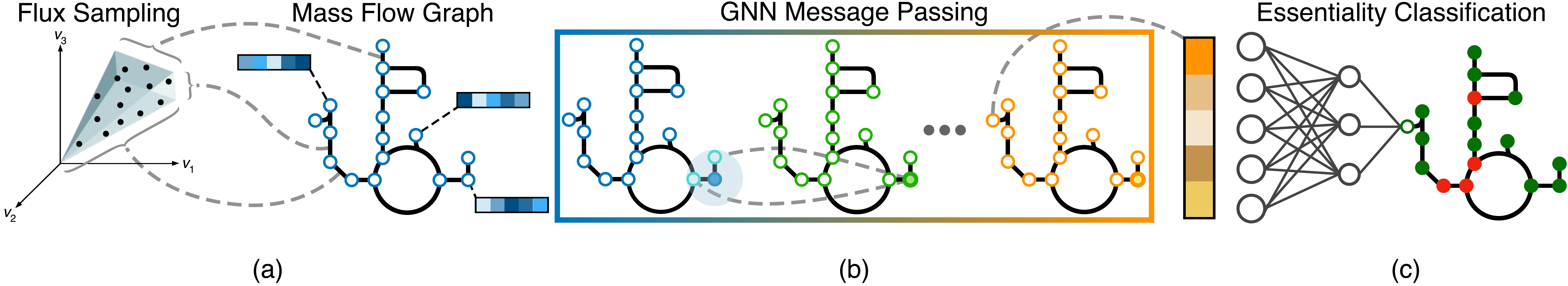}
    \caption{Architectural diagram of FluxGAT - where (a) flux sampling is used to construct a weighted graph of a metabolic reaction network. (b) The adjacency matrix, along with node features containing chemical properties of reactions, are then passed through an optimised graph attention network (GAT) which iteratively generates new representations for each reaction using neural message-passing layers. (c) Lastly, a dense neural layer and sigmoid activation function are used to perform binary classification from each node's final embedding, learning to predict if a reaction is essential (red) or non-essential (green) for cell growth.}
    \label{fig:architectural}
\end{figure*}

Gene essentiality is a fundamental concept in genomics, defined by the necessity of specific genes for the reproductive success of a cell or an organism \cite{rancati2018emerging}. Identification of essential genes is imperative because it provides an understanding of the basic requirements of a cell \cite{lachance2019minimal}, accelerates the discovery of drug targets \cite{cacheiro2020human}, and guides the engineering of organisms for chemical production \cite{niu2023high}.

Recent advancements in genome sequencing and CRISPR-based genome editing technologies have substantially advanced our ability to identify essential genes; however, laboratory methods are still time-consuming, costly, and only sometimes feasible across all organisms. The field has seen a significant shift towards \textit{in-silico} approaches for predicting gene essentiality, which analyse intrinsic genomic features, topological features of biological networks \cite{li2020network}, and combinations of features using machine learning (ML) \cite{aromolaran2021machine}.

ML approaches have gained recent traction due to their ability to complement experimental methods and generalise across organisms. These approaches involve training classifiers using known \textit{essential} and \textit{non-essential} genes from model organisms, integrating diverse categories of features such as gene and protein sequences, network topology, gene ontology and homology \cite{aromolaran2021machine,li2020network,freischem2022prediction}. While most of these approaches utilise traditional ML methods, such as decision trees (DTs) and support vector machines (SVMs), a smaller number employ deep learning-based neural networks. This choice is usually due to the trade-off between enhanced representation power versus the challenges with interpretability of \say{black-box} deep models.

Sequence-based ML methods, which utilise gene and protein sequences to determine gene essentiality, are commonly used but can be limited by the availability and quality of sequence data, especially for non-model organisms. Feature generation is often used to transform sequences into numerical representations; however, the lack of standard tools for certain features can introduce semantic errors in the analysis. Network-based methods, on the other hand, focus on interactions within biological networks and can predict essentiality from the inherent structural properties of cells without the need for sequencing data. These methods analyse the topological characteristics of nodes in biological networks, such as protein-protein interaction networks (PINs), to infer gene essentiality. The essentiality of a protein, for example, is closely associated with its topological characteristics within PINs \cite{dhasmana2020topological}. 

Graph neural networks (GNNs) represent a novel deep learning technique, particularly tailored for dealing with graph-structured data \cite{wu2022graph}. This property makes them well-suited for tasks where understanding the relationships and interactions within a network is crucial. Unlike the ML methods previously mentioned, which rely on the assumption of \textit{independent and identically distributed} (i.i.d.) data, GNNs can effectively capture and leverage the structure of nodes within a graph, breaking the i.i.d assumption, which enables them to generalise better in complex, interconnected systems \cite[p.~13]{hamilton2020graph}. GNNs, therefore, are likely to become central to single-cell analysis within biomedicine in the coming years \cite{lazaros2023graph}. In gene essentiality prediction, topology features are acknowledged for their strong discriminatory capabilities and less direct correlation with the biological functions of genes \cite{li2020network}, enhancing generalisability across organisms.

In metabolic engineering, an expanding field of synthetic biology, flux balance analysis (FBA) is arguably the most popular mathematical approach for studying cellular phenotypes and is commonly used to predict gene essentiality. FBA utilises constraint-based modelling (CBM) and linear programming to predict the distribution of metabolites across a metabolic network under steady-state conditions \cite{orth2010flux}. This method enables the analysis of the potential impact of gene deletions on the metabolic network, aiding in the identification of essential genes for organism survival under specific conditions. By using genome-scale metabolic models (GSMMs) of cells, FBA can predict the growth rate of an organism or the production rate of a metabolite, assuming a predefined cellular fitness objective. 

FBA has certain limitations, notably in the requirement to define a cellular objective function. This necessity arises from FBA's approach to modelling very large metabolic networks, which results in an underdetermined system with a near-infinite number of feasible solutions that satisfy the biological constraints \cite{wintermute2013objective}. The definition of an objective function, such as maximisation of biomass production or product yield, is essential for FBA to yield a specific solution. This reliance on a predefined objective introduces an element of observer bias, as the chosen objective may not fully encapsulate the actual biological goals of an organism \cite{schnitzer2022choice,ravi2021deltafba}, which are inherently dynamic and adapted based on the environment or genetic composition.

Flux sampling has been developed to address this limitation of FBA, negating the need for a predefined objective function by exploring a wide range of possible flux states that a metabolic network can adopt, given its constraints \cite{herrmann2019flux}. To achieve this, a sampling algorithm generates a sequence of feasible solutions until the entire solution space is analysed. Upon convergence, a probability distribution of steady-state reaction fluxes is generated that can be used for downstream analysis, such as studying metabolism under changing environmental conditions \cite{herrmann2019flux}. Flux sampling has demonstrated superior performance over FBA in predicting flux distributions \cite{strain2023reliable,gelbach2024flux}, particularly in complex mammalian cell lines where a single objective function is unlikely to be representative of nature. This highlights the need to develop objective-free methods for predicting gene essentiality in cells where objectives are less understood and more dynamic, ensuring a more general representation of their metabolic behaviours and responses.

In this work, we introduce FluxGAT, Figure \ref{fig:architectural}, a novel approach for metabolic gene essentiality prediction, focusing on its capability to learn meaningful representations from flux sampling data derived from the most complete GSMM of Chinese hamster ovary (CHO) cells \cite{yeo2020enzyme}, the most commonly used mammalian hosts in the biopharmaceutical industry \cite{yamano2020establishment}. Average flux values are derived from probability distributions and used to construct a graphical representation of chemical mass flow whereby nodes represent enzymatic reactions. The utilisation of GNNs and flux sampling data is particularly novel as it mitigates the inherent observer bias typically encountered in traditional methods that require the definition of a cellular objective. This innovation is crucial for mammalian and non-model organism where existing models often fall short. The following sections will outline the construction of a graphical representation of flux sampling data, the architectural design of FluxGAT, and will demonstrate its predictive accuracy through comprehensive evaluations, comparing it against traditional methods.

\section{Related Work}
\paragraph{ML and Genome-Scale Metabolic Models} 
ML-based methods have demonstrated comparable performance to FBA for predicting metabolic gene essentiality, using only wild-type flux distributions where cellular objectives are more likely to hold \cite{freischem2022prediction}. By only applying FBA to the wild-type cell, they eliminate the need to assume that deletion strains optimise the same fitness objective as the wild type. This work utilised four traditional ML classifiers to leverage manually extracted features from a graph-structured representation of metabolic fluxes, where nodes represent enzymatic reactions, and edges quantify the mass flow of metabolites between these reactions. Although the model could capture most of the essential genes determined experimentally, there was a high rate of false positive predictions.

\paragraph{GNNs and Genome-Scale Metabolic Models}
The same research group recently introduced an enhancement to their prior work by employing a GNN framework for increased predictive power over conventional ML methods for classifying essentiality using the same graphical representation. Comparative analysis of their model's predictions with FBA reveals a performance comparable to FBA in identifying \textit{essential} genes. However, it is noteworthy that FBA exhibits significantly superior performance in identifying \textit{non-essential} genes. The significance of their results lies in the generalisation ability of the model to accurately predict gene essentiality across various growth conditions. A critical limitation; however, is the reliance on the assumption of a wild-type fitness objective, which restricts the model's applicability to model organisms where such behaviour is well-documented. This work motivates the analysis of enzymatic reaction networks in predicting gene essentiality.

\paragraph{GNNs and Protein-Protein Interaction Networks}
A GNN model was employed to predict gene essentiality using PPI network data \cite{joo2020epgat}, significantly outperforming traditional ML and network-based methods. However, challenges arise due to the inherent noise and the prevalence of false positive connections in PPI networks. Additionally, the existence of multiple protein interaction databases, each presenting considerably varied network structures, can lead to inconsistent and unstable outcomes. This highlights the importance of GSMMs as highly utilised tools in systems biology, offering more robust frameworks for studying cellular phenotypes.

\paragraph{Flux Sampling and Gene Essentiality} To our knowledge, there is currently no existing work, utilising ML or other methodologies, that focuses on analysing GSMM flux sampling data for the specific aim of predicting gene essentiality. Consequently, this limits the direct comparability of our approach to existing methods, making FBA the primary benchmark for evaluating performance.

\section{Methodology}
\subsection{Task Definition}
In this study, we employ a GNN model to address the challenge of binary node label classification to determine the biological essentiality of reactions and their associated genes. Our approach is inherently a semi-supervised task as the GNN leverages the connectivity of the entire network, including test nodes whose features and labels are masked during training. The data utilised is a single graph, \( G = (V, E) \), which is a reconstruction of the CHO cell metabolic reaction network using flux sampling data, where \( V \) is the set of nodes (reactions) and \( E \) is the set of edges (interactions). The features used by our model, denoted as \( X \), represent the network topology with weighted edges alongside node features that encapsulate the chemical properties of the reactions. The binary node labels, \( Y \), being predicted, signify experimentally determined essentiality values. The graph's data structure, comprising a weighted and directed graph, reflects the intricacies of metabolic pathways, correlating the network topology and node features directly to the essentiality of reactions. This setup allows for a comprehensive understanding and prediction of gene essentiality.

\subsection{Flux Sampling}
\label{sec:sampling}
When modelling reaction networks within an organism at the genome scale, which often involves thousands of reactions and metabolites, the classical approach in systems biology of dynamical modelling becomes infeasible due to the time and resources required to obtain kinetic parameters experimentally. 

Constraint-based modelling is an approach to analysing such networks without detailed kinetic data. For a given reaction network, the relation between the $m$-metabolites and $n$-reactions is described in the $m$x$n$ stoichiometric matrix $S$. A positive stoichiometric coefficient $S_{i,j}$ means that metabolite $i$ is produced by reaction $j$, and a negative entry indicates that the metabolite is consumed in that reaction. The primary constraints limit the rate and directionality of each reaction in the network. Then, a steady-state assumption ensures that all metabolite concentrations stay constant over time (mass balance). The dynamics of the network at steady-state are defined as,

\begin{equation}
\frac{d\mathbf{x}}{dt} = S\mathbf{v} = 0, \hspace{5mm} \mathbf{v}^{lb} \leq \mathbf{v} \leq \mathbf{v}^{ub}. \label{eq:fba}
\end{equation}

\begin{figure}[t!]
    \centering
    \includegraphics[width=0.4\columnwidth]{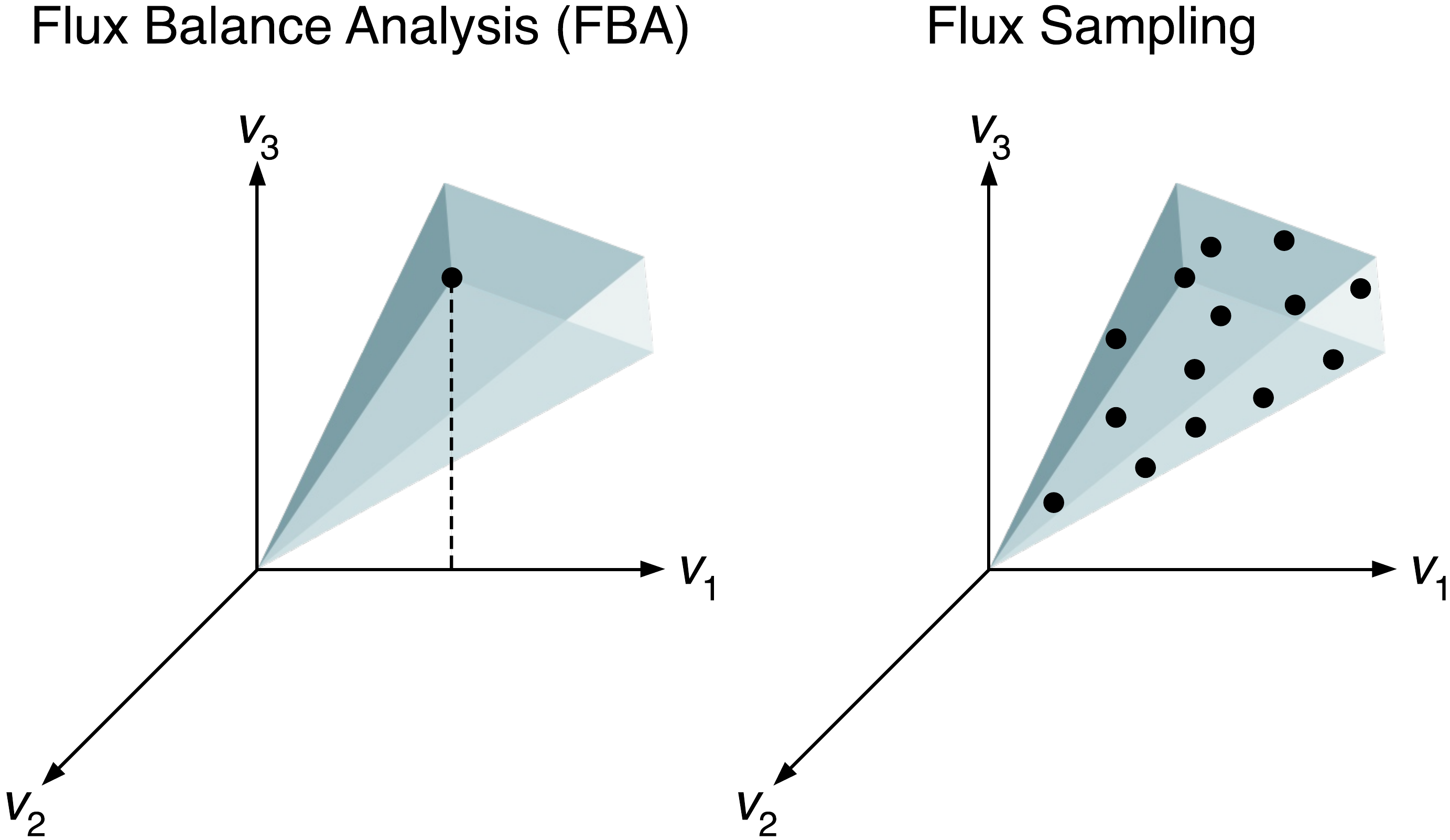}
    \caption{Flux balance analysis (FBA) generates a single non-unique solution, and flux sampling generates a probability distribution of feasible solutions of the constrained solution space (blue region) of reaction fluxes $\mathrm{v_1}
$, $\mathrm{v_2}$, $\mathrm{v_3}$.}
    \label{fig:flux_sampling}
\end{figure}

Where $\mathbf{x}$ is a vector of metabolite concentrations and $\mathbf{v}$ is a vector of flux rates, constrained by lower and upper bounds for each reaction. FBA introduces an objective function and uses linear programming to find a non-unique flux distribution $\textbf{v}^*$, which satisfies the constraints. Flux sampling, on the other hand, aims to uniformly sample the feasible solution space, given the constraints, avoiding the need to specify an objective function. Figure \ref{fig:flux_sampling} visualises this distinction.

Markov chain Monte Carlo (MCMC) methods are a class of algorithms used for sampling complex, high-dimensional probability distributions, making them well-suited for analysing the solution space of reaction networks. The Hit-and-Run (HR) algorithm is a deterministic MCMC technique that has emerged as the primary means for sampling large, genome-scale networks due to its efficiency and convergence time \cite{fallahi2020comparison}. In this work, an implementation of the artificial centring hit-and-run (ACHR) algorithm, known as \textit{optGpSampler} \cite{megchelenbrink2014optgpsampler}, was used due to its established application to GSMMs.

In the first iteration ($a=0$) of the ACHR algorithm, an arbitrary starting point $\textbf{v}^{(0)} \in P$, is selected within an $N$ dimensional convex set $P$, giving an initial centre of $\hat{\textbf{c}} = \textbf{v}^{(0)}$. The algorithm then generates a chain of samples by iteratively performing the following three steps. First, computing a direction to travel in $\boldsymbol{\theta}$,

\begin{equation}
    \boldsymbol{\theta}^{(a)} = \frac{\textbf{v}^{(i)} - \hat{\textbf{c}}}{\norm{\textbf{v}^{(a)} - \hat{\textbf{c}}}}, \quad \text{where $i \sim U\{0, 1, \ldots, a\}.$}
\end{equation}

Then, after randomly generating a step size $\lambda^{(a)}$, a new sample is calculated $\textbf{v}^{(a+1)} = \textbf{v}^{(a)} + \lambda^{(a)} \boldsymbol{\theta}^{(a)}$. Lastly, updating the artificial centre by setting,

\begin{equation}
    \hat{\textbf{c}} = \frac{a \hat{\textbf{c}} + \textbf{v}^{(a)}}{a + 1}.
\end{equation}

This process is repeated until convergence, as confirmed by diagnostic tests, resulting in probability distributions of possible flux values for each reaction. Then, a single flux distribution is calculated by averaging the probability distribution for each reaction to generate a one-dimensional flux vector $\Bar{\textbf{v}}$, facilitating downstream analysis, as was done in \cite{strain2023reliable}.

\subsection{Graph Construction}
\label{sec:graph_construction}
There are numerous ways to construct a graphical representation of a given metabolic reaction network using its stoichiometry; however, many fail to capture the directionality of reactions and are dominated by pool metabolites that appear in many reactions, such as water. The mass flow graph (MFG) was designed to address these common limitations by utilising flux distributions to encode the directionality of metabolic flows whilst mitigating the over-representation of pool metabolites \cite{beguerisse2018flux}, producing a directed graph. The MFG also exhibits systemic changes to its topological and community structure under environmental and genetic perturbations, revealing redundancies and core structures within a reaction network. It is, therefore, well-suited for highlighting essential components.

An MFG can be constructed from a stoichiometry matrix $S$ whereby nodes represent reactions and a directed edge connects two nodes if the source reaction produces a metabolite consumed by the target reaction. As detailed in \cite{beguerisse2018flux}, starting with flux vector, obtained either through FBA $\textbf{v}^*$, or in our case, averaged across flux sampling iterations $\Bar{\textbf{v}}$, we first unfold the vector into the forward and reverse components of each reaction,

\begin{equation}
{\Bar{\mathbf{v}}}_{2m} = \left[ {\begin{array}{*{20}{c}} {\Bar{\mathbf{v}}}^+ \cr {\Bar{\mathbf{v}}^-} \end{array}} \right] = \frac{1}{2}\left[ {\begin{array}{*{20}{c}} {{\mathrm{abs}}\left( {\Bar{\mathbf{v}}} \right) + {\Bar{\mathbf{v}}} } \cr {{\mathrm{abs}}\left( {\Bar{\mathbf{v}} } \right) - {\Bar{\mathbf{v}}} } \end{array}} \right].
\end{equation}

Then, we can define the modified stoichiometry matrix as,

\begin{equation}
{\mathbf{S}}_{2m} = \left[ {\begin{array}{*{20}{c}} {\mathbf{S}} & { - {\mathbf{S}}} \end{array}} \right]\left[ {\begin{array}{*{20}{c}} {{\mathbf{I}}_m} & 0 \cr 0 & {{\mathrm{diag}}\left( {\mathbf{r}} \right)} \end{array}} \right],
\end{equation}

where $r$ is an $m$-dimensional Boolean reversibility vector such that $r_j=1$ if reaction $j$ is reversible and $0$ otherwise. Lastly, the adjacency matrix of the MFG is given by,

\begin{equation}
{\mathbf{M}}\left( {{\Bar{\mathbf{v}}} } \right) = \left( {{\mathbf{S}}_{2m}^ + {\Bar{\mathbf{V}}}} \right)^T {\mathbf{J}}_v^\dagger \left( {{\mathbf{S}}_{2m}^ - {\Bar{\mathbf{V}}} } \right),
\end{equation}

where ${\Bar{\mathbf{V}}}$ and ${\bf{J}}_v$ are defined as ${\mathrm{diag}}\left( {{\Bar{\mathbf{v}}}_{2m}} \right)$ and ${\mathrm{diag}}\left( {{\mathbf{S}}_{2m}^ +} {{\Bar{\mathbf{v}}}_{2m}} \right)$ respectively, and $\dagger$ denotes the matrix pseudoinverse, with,

\begin{equation}
    \begin{array}{*{20}{l}} {{\mathrm{Production:}}\quad } \hfill & {{\mathbf{S}}_{2m}^ + = \frac{1}{2}\left( {{\mathrm{abs}}\left( {{\mathbf{S}}_{2m}} \right) + {\mathbf{S}}_{2m}} \right)}, \hfill \cr {{\mathrm{Consumption:}}\quad } \hfill & {{\mathbf{S}}_{2m}^ - = \frac{1}{2}\left( {{\mathrm{abs}}\left( {{\mathbf{S}}_{2m}} \right) - {\mathbf{S}}_{2m}} \right).} \hfill \end{array}
\end{equation}


\subsection{Node Label Generation}
Considering that the MFG comprises a set of reactions as nodes rather than a set of genes, we must initially map the gene essentialities onto the corresponding set of reaction essentialities to train and test a classifier. To perform this mapping, the method outlined in \cite{ponce2020inconsistent} is utilised, where the authors address the inconsistent treatment of gene protein reaction (GPR) rules by defining a standard approach to handling them. GPR rules describe how genes relate to protein complexes and the reactions they catalyse. They contain Boolean logic operators that allow us to classify reactions as active/inactive using gene expression.

These GPR rules can also be processed to translate gene essentiality into node essentiality labels according to the following function $f$, applied to each reaction,

\begin{equation}
\begin{split}
f(\oplus, g_1, \ldots, g_n) = 
\begin{cases} 
\max(g_1, \ldots, g_n) & \text{if only ANDs} \\
\min(g_1, \ldots, g_n) & \text{if only ORs} \\
\text{\parbox{3cm}{Resolve ANDs \\ then ORs}} & \text{if mixed}
\end{cases}
\end{split}
\label{eq:gpr}
\end{equation}

where,

\begin{itemize}
    \item $\oplus$: Represents the logical operator or the combination of operators within the GPR rule for a given reaction.
    \item $g_1, \ldots, g_n$: The sequence of gene essentiality values (1 for essential and 0 for non-essential).
    \item \textit{Resolve ANDs then ORs}: For mixed operators, standard logical precedence rules apply (AND operations are typically evaluated before OR operations unless parentheses indicate otherwise).
\end{itemize}

For example, in the case of a reaction with the GPR rule and gene essentiality labels \say{1 \& (0 $\vert$ 1) \& 0}, the operation within parentheses (OR operation) would typically be evaluated after resolving all AND operations. However, given the explicit grouping by parentheses, the OR operation is resolved first in this segment, followed by the AND operation. Here, the reaction would be considered essential (1) due to the first gene being essential.

\subsection{Node Feature Generation}
\label{sec:features}
GNNs exhibit state-of-the-art performance across various network-based tasks, primarily due to their ability to combine representations of network topology with \textit{side information} such as node features. This section outlines the node (reaction) feature generation process to create embeddings that complement the adjacency matrix, $M$.

Our generated node features contain biologically meaningful information for the downstream classification task of essentiality prediction. This approach involved compiling a list of the reactants and products of each reaction, representing additional chemical properties not captured within the MFG. These lists encode each reaction's starting material and end result in a structured format, emphasising relationships between nodes.

A one-hot encoding was employed to transform node features into a numerical format suitable for a GNN. To achieve this, we let \( R \) be the set of all reactants and products across reactions, with \( |R| \) being its cardinality. For each reaction \( i \), a one-hot encoded vector \( \mathbf{v}_i \) is constructed, where \( \mathbf{v}_i \in \{0, 1\}^{|R|} \). For each reactant or product \( j \) in \( R \), the corresponding element in \( \mathbf{v}_i \), denoted as \( v_{ij} \), is defined as:

\[
v_{ij} = 
\begin{cases} 
    1 & \text{if reactant/product } j \text{ is involved in reaction } i, \\
    0 & \text{otherwise.}
\end{cases}
\]

\subsection{Graph Representation Learning}
\label{sec:GNN}
A GNN model is used to perform representation learning of the MFG due to their unique ability to learn representations that combine features of individual reactions with the patterns of interactions and relationships that define the reaction network topology. Specifically, we use a graph attention network (GAT) \cite{velivckovic2017graph} due to its integration of self-attention mechanisms, which allow for a learnt weighting of neighbouring nodes based on their relevance to a given reaction in the context of the representation learning task. Such mechanisms also enhance the interpretability of the model, providing insights into the prioritisation of specific nodes (reactions).

More formally, GNNs contain \textit{neural message passing} layers in which vector messages are exchanged between nodes and updated using neural networks. For each message-passing layer within a GNN, a \textit{hidden embedding} $\mathbf{h}_u^{(i)}$ corresponding to each node $u \in \mathcal{V}$ is updated by aggregated information from $u$'s graph neighbourhood $\mathcal{N}(u)$. Each message-passing update iteration can be defined as follows,

\begin{subequations}
\begin{align}
\textbf{h}_u^{(i+1)} &= \text{UPDATE}^{(i)} \left( \textbf{h}_u^{(i)}, \right. \nonumber \\
&\quad \left. \text{AGGREGATE}^{(i)} ( \{ \textbf{h}_v^{(i)} \, \forall v \in \mathcal{N}(u)\} ) \right) \\
&= \text{UPDATE}^{(i)} \left( \textbf{h}_u^{(i)}, \textbf{m}_{\mathcal{N}(u)}^{(i)} \right),
\end{align}
\end{subequations}

where UPDATE and AGGREGATE are arbitrary differentiable functions and $\textbf{m}_{\mathcal{N}(u)}$  is the \say{message} that is aggregated from $u$'s one-hop graph neighbourhood $\mathcal{N}(u)$ at iteration $i$. Here, we use a modified version of the original GAT, assigning a learnable attention weight to each neighbour based on node and edge features. This attention weight determines the \textit{importance} of nodes during the aggregation step, with each message defined as,

\begin{equation}
    \textbf{m}_{\mathcal{N}(u)} = \sum_{v \in \mathcal{N}(u)} \alpha_{u,v} \mathbf{h}_v,
\end{equation}

where $\alpha_{u,v}$ denotes the attention given to neighbour $v \in \mathcal{N}(u)$ during aggregation at node $u$. The attention weights are calculated based on the importance of the features of the source node, the target node, and the edge, as follows,

\begin{equation}
\alpha_{u,v} = \frac{\exp \left( \textbf{a}^{\top} \left[ \textbf{W}\textbf{h}_{u} \oplus \textbf{W}\textbf{h}_{v} \oplus \textbf{W}\textbf{e}_{u,v} \right] \right)}{\sum_{v' \in \mathcal{N}_(u)} \exp \left( \textbf{a}^{\top} \left[ \textbf{W}\textbf{h}_{u} \oplus \textbf{W}\textbf{h}_{v'} \oplus \textbf{W}\textbf{e}_{u,v'} \right] \right)},
\end{equation}

where $\mathbf{a}$ is a trainable attention vector, $\mathbf{W}$ are trainable matrices unique to each feature, and $\oplus$ denotes the concatenation operation. Multi-dimensional edge features between nodes $u$ and $v$ are represented by $\textbf{e}_{u,v'}$, which, in our case, are equal to scalar edge weights of the MFG representing chemical mass flow between reactions.

The initial node embeddings at $i=0$, $\mathbf{h}_u^{(0)}$, are set to the one-hot encodings detailed in Section \ref{sec:features}. After running $N$ iterations of the GNN message passing, we can use the output of the final layer to define the embeddings for each node, i.e.,

\begin{equation}
    \mathbf{z}_u = \mathbf{h}_u^{(N)}, \forall u \in \mathcal{V}.
\end{equation}

Then, to perform node classification, $\mathbf{z}_u$ is passed through a dense neural layer to generate a single logit $z_u$, which is passed through a sigmoid activation function to determine the binary class (essentiality) for node (reaction) $u$.

\section{Experiment}
\subsection{Dataset}
The artificial centring hit-and-run (ACHR) algorithm, detailed in Section \ref{sec:sampling}, was applied \footnote{The source code for this project is available on GitHub: \url{https://github.com/kierensharma/FluxGAT}.} to the wild-type iCHO2291 GSMM of CHO cells \cite{yeo2020enzyme}, which contains $m=3,972$ metabolites and $n=6,236$ reactions. Sampling was performed for 50,000 iterations with a \textit{thinning} value of 1000 to reduce the memory footprint of the sampling results whilst still ensuring meaningful convergence \cite{galuzzi2022best}. Averaging the probability distributions for each reaction resulted in a single flux vector $\Bar{\textbf{v}}$ with 5,316 non-zero reaction fluxes. According to the methods in Section \ref{sec:graph_construction}, an MFG was constructed containing 4,733 nodes. The MFG also contains 335,024 weighted edges, representing the directionality and strength of metabolite flow between reactions. Function $f$ (Equation \ref{eq:gpr}) was applied to generate binary essentiality labels for nodes in the MFG using experimental gene essentiality values taken from a genome-wide CRISPR knockout screen of CHO cells \cite{xiong2021optimized}. This study identified 1,980 genes out of the 15,028 targeted which negatively affect cell proliferation and are therefore deemed essential. The iCHO2291 model contains 2,291 functionally modelled genes representing CHO cell metabolism, 392 of which are essential according to \cite{xiong2021optimized}. Applying $f$ to all reactions within the model containing GPR rules (4,182), classified 3,296 reactions as non-essential and 886 as essential. These reaction essentiality values allowed us to generate node labels for 3,310 nodes within the MFG, with 2,498 being non-essential (0) and 812 essential (1).

\subsection{Model Architecture and Training}
\begin{figure*}
    \centering
    \begin{subfigure}[c]{0.6\textwidth}
        \includegraphics[width=\textwidth]{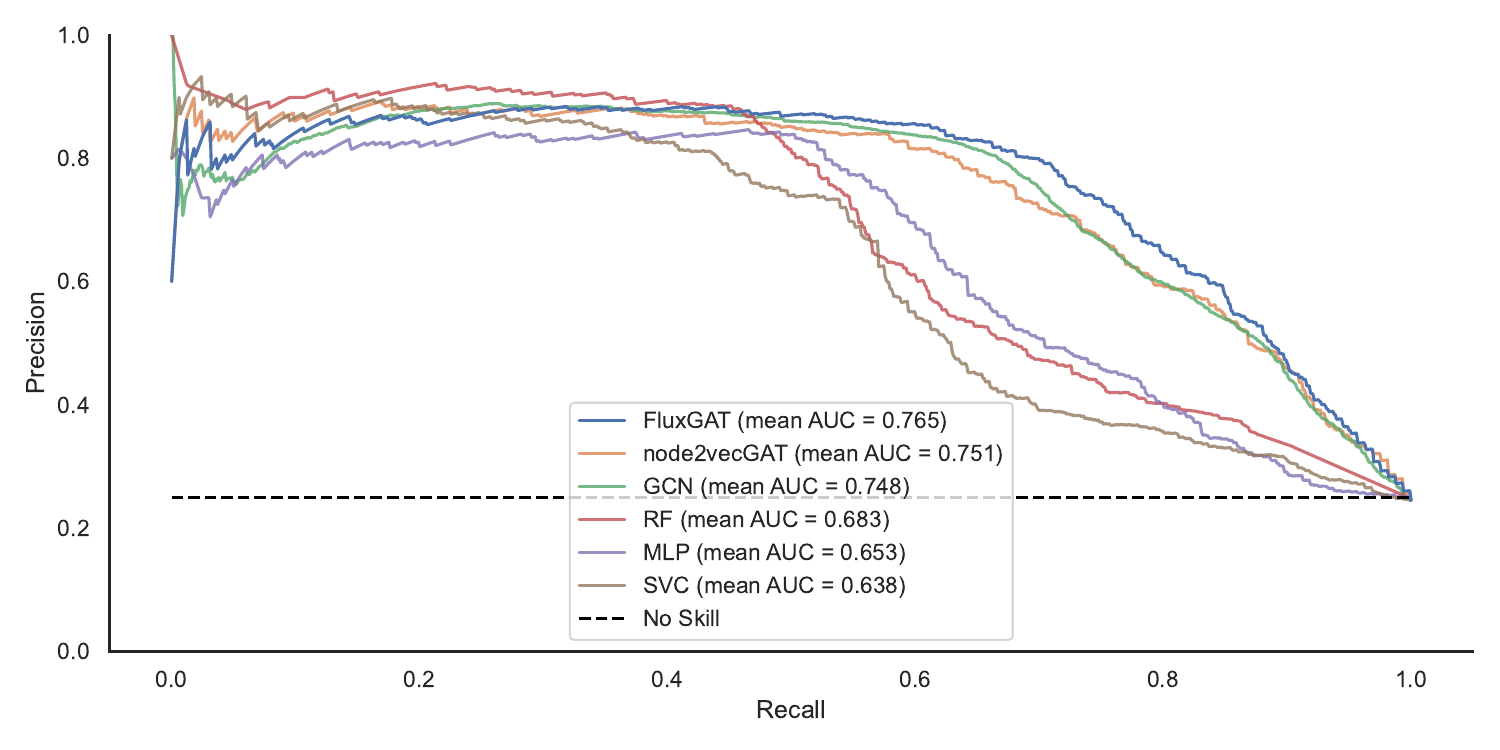}
        \caption{}
        \label{fig:PRAUC}
    \end{subfigure}
    \hspace{0.5cm} 
    \begin{subfigure}[c]{0.35\textwidth}
        \includegraphics[width=\textwidth]{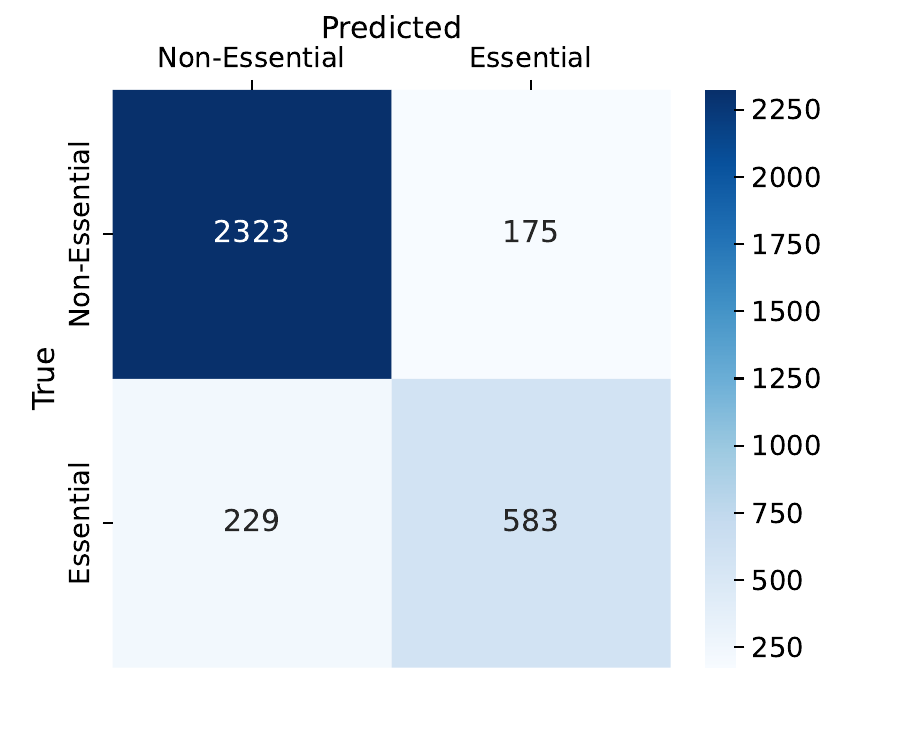}
        \caption{}
        \label{fig:conf}
    \end{subfigure}
    \caption{(a) Precision-recall (PR) curves for FluxGAT, node2vecGAT, GCN and three standard binary classifiers, averaged across 25 evaluations (5 repetitions of 5-fold cross-validation). The dashed line represents the precision (0.249) of the no-skill classifier given the class imbalance. (b) Confusion matrix showing the average classification performance of FluxGAT on the 3,310 experimentally determined reaction essentiality labels.}
\end{figure*}

FluxGAT utilises a supervised learning approach to discern and comprehend complex patterns within node features and relationships within the MFG, as can be seen in Figure \ref{fig:embeddings}. This training method enabled the model to learn biologically meaningful representations to predict essentiality labels. During evaluation, the graph was partitioned into training (80\%) and testing (20\%) sets, ensuring node features designated for testing were masked during the training phase.

We employed k-fold cross-validation to partition the graph systematically into five distinct folds for training (2,648 nodes) and testing (662 nodes), ensuring each fold served as a testing set once. This procedure allowed every reaction in the graph to be part of both the training and testing sets across different iterations, mitigating potential biases and testing the generalisability of the model. FluxGAT contains an initial embedding layer to transform the sparse one-hot gene vectors into a lower-dimensional, continuous space, facilitating more efficient and meaningful feature processing in the subsequent GAT message-passing layers.

To fine-tune the performance of FluxGAT, a grid-search method was employed for hyperparameter optimisation of the underlying GAT architecture. The optimal architecture (Table \ref{tab:hyperparameters}) is initialised with 150-dimensional embeddings to transform 3,972-dimensional one-hot encodings into dense vectors. Two GAT message-passing layers, containing multi-head attention (2 heads) and 150 hidden channels, transform the dense node embeddings into a final representation, which is fed to a fully connected layer to consolidate the features into a single logit per node, setting the stage for binary classification using a sigmoid activation function with an optimised threshold of 0.65. A weighted loss function helps to address the class imbalance in essentiality labels within the MFG to ensure a more balanced treatment of each class. The weight was determined based on the ratio of non-essential to essential reactions, precisely the number of zeros (2,498) to the number of ones (812). For training FluxGAT, the AdamW optimiser was employed to improve generalisation performance \cite{loshchilov2017decoupled}, complemented by regularisation strategies to prevent overfitting, which included weight decay, dropout and an early-stopping procedure.

\subsection{FluxGAT Performance}
\begin{table}[]
\centering
\begin{tabular}{lrrrr}
\hline
Method      & Accuracy                   & Precision                  & Recall                     & F1 Score                   \\ \hline
FluxGAT     & $\textbf{0.871} \pm \textbf{0.012}$ & $0.769 \pm 0.030$          & $\textbf{0.718} \pm \textbf{0.037}$ & $\textbf{0.743} \pm \textbf{0.023}$ \\
node2vecGAT & $0.864 \pm 0.014$          & $0.741 \pm 0.040$          & $0.693 \pm 0.038$          & $0.714 \pm 0.023$          \\
GCN         & $0.870 \pm 0.013$          & $0.754 \pm 0.030$          & $0.699 \pm 0.032$          & $0.725 \pm 0.030$          \\
RF          & $0.848 \pm 0.010$          & $0.786 \pm 0.040$          & $0.520 \pm 0.041$          & $0.625 \pm 0.039$          \\
MLP         & $0.854 \pm 0.011$          & $0.813 \pm 0.040$          & $0.528 \pm 0.028$          & $0.639 \pm 0.027$          \\
SVC         & $0.827 \pm 0.010$          & $\textbf{0.843} \pm \textbf{0.050}$ & $0.366 \pm 0.027$          & $0.509 \pm 0.025$          \\ \hline
\end{tabular}%
\vspace{2mm}
\caption{\textbf{Comparative Performance of FluxGAT and Benchmark Models.} A detailed overview of the mean and standard deviation for test accuracy, precision, recall, and F1 score, aggregated over 25 evaluations (comprising five repetitions of 5-fold cross-validation) for each binary classifier.}
\label{tab:performance}
\end{table}

We begin by comparing FluxGAT against three standard binary classifiers: support vector classifier (SVC), multilayer perceptron (MLP), and random forest (RF). The performance of each model, including accuracy, precision, recall, and F1 score, was averaged across five repetitions of 5-fold cross-validation to ensure robustness and consistency in our evaluation.

A manual feature extraction process was employed to generate node features from the MFG to implement the standard binary classifiers. This process involved creating vectors for each node, composed of the incoming and outgoing edge weights associated with each node, effectively capturing the flow of metabolites into and out of each metabolic reaction's one-hop neighbourhoods.

Figure \ref{fig:PRAUC} presents the precision-recall (PR) curves for FluxGAT and the baseline classifiers, averaged across the 25 evaluations. This choice of PR curves was due to the class imbalance in the MFG, as they provide a more informative performance measure under such conditions compared to traditional receiver operating characteristic (ROC) curves. FluxGAT achieves the highest precision-recall area under the curve (PR-AUC) at 0.765, a significant improvement over the \textit{no-skill} classifier, which has a precision of 0.249, indicative of the class imbalance.

In addition to comparing FluxGAT with standard binary classifiers, we further evaluated its performance against the same GAT architecture but implemented with \textit{node2vec} embeddings \cite{grover2016node2vec}. This comparison allows us to determine the impact of the custom embeddings on FluxGAT's performance. The node2vecGAT model serves as a GNN benchmark, as it represents a well-established method for generating node embeddings based on graph structure but without the domain-specific feature engineering employed in FluxGAT. Lastly, we evaluated the performance of our framework by substituting a graph convolutional network (GCN) for the GAT architecture. Both GNN benchmarks surpassed the standard binary classifiers, underscoring the efficacy of the message-passing framework detailed in Section \ref{sec:GNN} in surpassing manual feature extraction for creating topological representations of nodes. Notably, FluxGAT, with its incorporation of attention mechanisms and custom node features, achieved a significant enhancement in Precision-Recall Area Under the Curve (PR-AUC). Table \ref{tab:performance} compares FluxGAT's performance to baseline models.

\subsection{Comparison with Flux Balance Analysis}
\begin{figure}[t!]
    \centering
    \begin{subfigure}[b]{0.3\columnwidth}
        \includegraphics[width=\columnwidth]{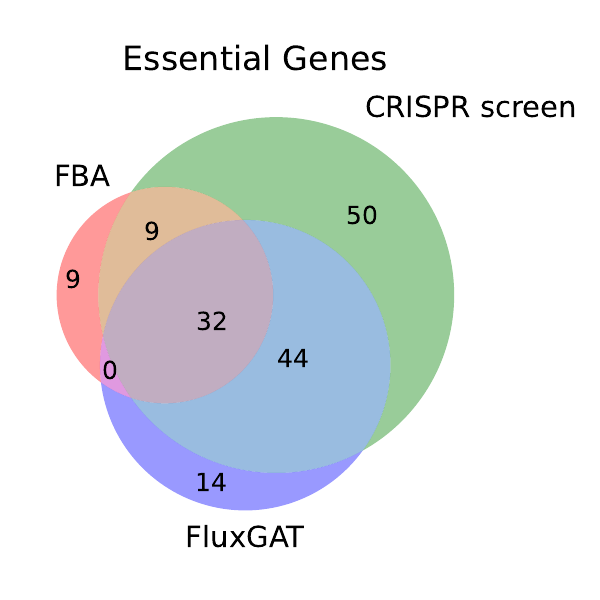}
        \caption{}
        \label{fig:ess}
    \end{subfigure}
    \hspace{0.5cm} 
    \begin{subfigure}[b]{0.3\columnwidth}
        \includegraphics[width=\columnwidth]{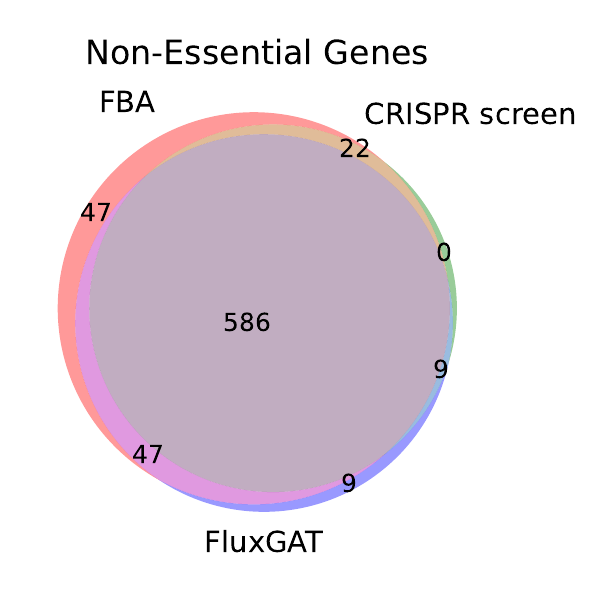}
        \caption{}
        \label{fig:noness}
    \end{subfigure}
    \caption{Comparison of gene essentiality predictions with experimental CRISPR screen, for genes involved in MFG reactions with a unique inverse mapping. (a) Essential gene predictions by FBA and FluxGAT. (b) Non-essential gene predictions by the same methods.}
    \label{fig:gene_predictions}
\end{figure}

This section details converting FluxGAT's binary reaction predictions into gene essentiality labels. To perform this conversion, we applied the inverse function of $f$ (referenced in Equation \ref{eq:gpr}) to the reactions where GPR rules allow a unique inverse mapping. Specifically, this includes reactions with one-to-one mappings to genes, essential reactions governed only by OR operators, and non-essential reactions exclusively involving AND operators in their GPR rules. Out of the 2,085 genes present in the MFG, 752 are involved in uniquely invertible mappings.

Applying this inverse mapping to reactions correctly labelled by FluxGAT across five repetitions, we can compare the essentiality classification of genes to those made by FBA (also across five repetitions). Figure \ref{fig:gene_predictions} shows the agreement between FluxGAT and FBA gene essentiality labels with those determined experimentally by a genome-wide CRISPR screen \cite{xiong2021optimized}. FluxGAT can identify 32 of the 41 essential genes correctly labelled by FBA and an additional 44 genes missed by FBA. However, this comes at the cost of a further 14 false-positive predictions. FluxGAT can also identify nearly all non-essential genes correctly labelled by FBA but with only 56 false negatives instead of 94.

As observed by Strain et al. \cite{strain2023reliable}, the iCHO2291 GSMM displays very high specificity, rarely classifying a non‐essential gene as essential, but shows lower sensitivity, failing to capture all essential genes when using FBA. We also observed this in our analysis when classifying the 752 genes in the MFG, whereby FBA displays a specificity of 0.985 but a sensitivity of 0.304. FluxGAT shows similar specificity, with 0.977, but an increased sensitivity of 0.576, capturing nearly double the number of essential genes as FBA.

\section{Discussion}
In this research, we have introduced FluxGAT, a proof-of-concept GNN-based approach for predicting metabolic gene essentiality without the inherent observer bias of FBA, the most commonly used method within systems biology. The application of FluxGAT to Chinese hamster ovary cells demonstrates a significant advancement in accurately predicting gene essentiality by learning from flux sampling distributions and metabolic reaction network topology.

The core innovation of FluxGAT lies in its unique methodology, which combines data derived from flux sampling with a graph attention network framework. This approach enables the prediction of gene essentiality based on intrinsic characteristics of the cell's structure, eliminating the need for predefined cellular objectives. This feature is particularly advantageous for studying mammalian and non-model organisms, where objectives aren't well-defined, offering an unbiased approach to predicting cellular phenotypes.

Our findings demonstrate that FluxGAT surpasses FBA in terms of sensitivity when applied to the most widely used mammalian cell line, a critical enhancement for various applications ranging from identifying drug targets to improving outcomes in metabolic engineering. However, this approach has limitations, as the model's complexity and the need for extensive computational resources may pose challenges for broader applications. Furthermore, the flux sampling method used can yield samples with thermodynamically infeasible loops, which, while part of the feasible solution space, do not align with biological reality. This aspect potentially affects FluxGAT's performance by including biologically irrelevant solutions. Sampling non-convex, loopless regions to avoid such loops presents a significant challenge due to the absence of standardised tools for this purpose. However, this limitation offers a direction for future work, suggesting the exploration of alternative sampling algorithms once they become available.

A crucial direction for our ongoing research is to assess FluxGAT's performance across various cell types and environmental conditions, to verify its generalisation abilities. This presents challenges due to the scarcity of genome-wide growth screens and the significant computational resources needed to analyse extensive reaction networks. We are actively exploring solutions to these obstacles to broaden the applicability and robustness of FluxGAT. Another promising research direction is the examination of the attention weights FluxGAT learns when predicting the essentiality of nodes. By interpreting these weights, we aim to uncover the biological significance of each node within the context of the entire reaction network.

In conclusion, FluxGAT's ability to predict gene essentiality directly from genotype, free from observer bias, demonstrates a method for more general gene essentiality prediction across diverse biological systems. This advancement highlights the significant potential of deep learning techniques in enhancing commonly used tools within systems biology. FluxGAT, therefore, represents a substantial stride towards advancing \textit{in-silico} methods used within personalised medicine and targeted drug development.

\bibliographystyle{unsrt}

\begin{thebibliography}{10}

\bibitem{rancati2018emerging}
Giulia Rancati, Jason Moffat, Athanasios Typas, and Norman Pavelka.
\newblock Emerging and evolving concepts in gene essentiality.
\newblock {\em Nature Reviews Genetics}, 19(1):34--49, 2018.

\bibitem{lachance2019minimal}
Jean-Christophe Lachance, S{\'e}bastien Rodrigue, and Bernhard~O Palsson.
\newblock Minimal cells, maximal knowledge.
\newblock {\em Elife}, 8:e45379, 2019.

\bibitem{cacheiro2020human}
Pilar Cacheiro, Violeta Mu{\~n}oz-Fuentes, Stephen~A Murray, Mary~E Dickinson,
  Maja Bucan, Lauryl~MJ Nutter, Kevin~A Peterson, Hamed Haselimashhadi, Ann~M
  Flenniken, Hugh Morgan, et~al.
\newblock Human and mouse essentiality screens as a resource for disease gene
  discovery.
\newblock {\em Nature communications}, 11(1):655, 2020.

\bibitem{niu2023high}
Kun Niu, Qiang Fu, Zi-Long Mei, Li-Rong Ge, An-Qi Guan, Zhi-Qiang Liu, and
  Yu-Guo Zheng.
\newblock High-level production of l-methionine by dynamic deregulation of
  metabolism with engineered nonauxotroph escherichia coli.
\newblock {\em ACS Synthetic Biology}, 12(2):492--501, 2023.

\bibitem{li2020network}
Xingyi Li, Wenkai Li, Min Zeng, Ruiqing Zheng, and Min Li.
\newblock Network-based methods for predicting essential genes or proteins: a
  survey.
\newblock {\em Briefings in bioinformatics}, 21(2):566--583, 2020.

\bibitem{aromolaran2021machine}
Olufemi Aromolaran, Damilare Aromolaran, Itunuoluwa Isewon, and Jelili Oyelade.
\newblock Machine learning approach to gene essentiality prediction: a review.
\newblock {\em Briefings in bioinformatics}, 22(5):bbab128, 2021.

\bibitem{freischem2022prediction}
Lilli~J Freischem, Mauricio Barahona, and Diego~A Oyarz{\'u}n.
\newblock Prediction of gene essentiality using machine learning and
  genome-scale metabolic models.
\newblock {\em IFAC-PapersOnLine}, 55(23):13--18, 2022.

\bibitem{dhasmana2020topological}
Anupam Dhasmana, Swati Uniyal, Anukriti, Vivek~Kumar Kashyap, Pallavi
  Somvanshi, Meenu Gupta, Uma Bhardwaj, Meena Jaggi, Murali~M Yallapu, Shafiul
  Haque, et~al.
\newblock Topological and system-level protein interaction network (pin)
  analyses to deduce molecular mechanism of curcumin.
\newblock {\em Scientific reports}, 10(1):12045, 2020.

\bibitem{wu2022graph}
Lingfei Wu, Peng Cui, Jian Pei, Liang Zhao, and Xiaojie Guo.
\newblock Graph neural networks: foundation, frontiers and applications.
\newblock In {\em Proceedings of the 28th ACM SIGKDD Conference on Knowledge
  Discovery and Data Mining}, pages 4840--4841, 2022.

\bibitem{hamilton2020graph}
William~L Hamilton.
\newblock {\em Graph representation learning}.
\newblock Morgan \& Claypool Publishers, 2020.

\bibitem{lazaros2023graph}
Konstantinos Lazaros, Dimitris~E Koumadorakis, Panagiotis Vlamos, and
  Aristidis~G Vrahatis.
\newblock Graph neural network approaches for single-cell data: A recent
  overview.
\newblock {\em arXiv preprint arXiv:2310.09561}, 2023.

\bibitem{orth2010flux}
Jeffrey~D Orth, Ines Thiele, and Bernhard~{\O} Palsson.
\newblock What is flux balance analysis?
\newblock {\em Nature biotechnology}, 28(3):245--248, 2010.

\bibitem{wintermute2013objective}
Edwin~H Wintermute, Tami~D Lieberman, and Pamela~A Silver.
\newblock An objective function exploiting suboptimal solutions in metabolic
  networks.
\newblock {\em BMC systems biology}, 7:1--16, 2013.

\bibitem{schnitzer2022choice}
Barbara Schnitzer, Linnea {\"O}sterberg, and Marija Cvijovic.
\newblock The choice of the objective function in flux balance analysis is
  crucial for predicting replicative lifespans in yeast.
\newblock {\em Plos one}, 17(10):e0276112, 2022.

\bibitem{ravi2021deltafba}
Sudharshan Ravi and Rudiyanto Gunawan.
\newblock $\delta$fba—predicting metabolic flux alterations using
  genome-scale metabolic models and differential transcriptomic data.
\newblock {\em PLoS Computational Biology}, 17(11):e1009589, 2021.

\bibitem{herrmann2019flux}
Helena~A Herrmann, Beth~C Dyson, Lucy Vass, Giles~N Johnson, and Jean-Marc
  Schwartz.
\newblock Flux sampling is a powerful tool to study metabolism under changing
  environmental conditions.
\newblock {\em NPJ systems biology and applications}, 5(1):32, 2019.

\bibitem{strain2023reliable}
Benjamin Strain, James Morrissey, Athanasios Antonakoudis, and Cleo Kontoravdi.
\newblock How reliable are chinese hamster ovary (cho) cell genome-scale
  metabolic models?
\newblock {\em Biotechnology and Bioengineering}, 2023.

\bibitem{gelbach2024flux}
Patrick~E Gelbach, Handan Cetin, and Stacey~D Finley.
\newblock Flux sampling in genome-scale metabolic modeling of microbial
  communities.
\newblock {\em BMC bioinformatics}, 25(1):45, 2024.

\bibitem{yeo2020enzyme}
Hock~Chuan Yeo, Jongkwang Hong, Meiyappan Lakshmanan, and Dong-Yup Lee.
\newblock Enzyme capacity-based genome scale modelling of cho cells.
\newblock {\em Metabolic engineering}, 60:138--147, 2020.

\bibitem{yamano2020establishment}
Noriko Yamano-Adachi, Rintaro Arishima, Sukwattananipaat Puriwat, and Takeshi
  Omasa.
\newblock Establishment of fast-growing serum-free immortalised cells from
  chinese hamster lung tissues for biopharmaceutical production.
\newblock {\em Scientific Reports}, 10(1):17612, 2020.

\bibitem{joo2020epgat}
Anderson~Tavares Joo~Schapke and Mariana Recamonde-Mendoza.
\newblock Epgat: Gene essentiality prediction with graph attention networks.
\newblock {\em arXiv preprint arXiv:2007.09671}, 2020.

\bibitem{fallahi2020comparison}
Shirin Fallahi, Hans~J Skaug, and Guttorm Alendal.
\newblock A comparison of monte carlo sampling methods for metabolic network
  models.
\newblock {\em Plos one}, 15(7):e0235393, 2020.

\bibitem{megchelenbrink2014optgpsampler}
Wout Megchelenbrink, Martijn Huynen, and Elena Marchiori.
\newblock optgpsampler: an improved tool for uniformly sampling the
  solution-space of genome-scale metabolic networks.
\newblock {\em PloS one}, 9(2):e86587, 2014.

\bibitem{beguerisse2018flux}
Mariano Beguerisse-D{\'\i}az, Gabriel Bosque, Diego Oyarz{\'u}n, Jes{\'u}s
  Pic{\'o}, and Mauricio Barahona.
\newblock Flux-dependent graphs for metabolic networks.
\newblock {\em NPJ systems biology and applications}, 4(1):32, 2018.

\bibitem{ponce2020inconsistent}
Miguel Ponce-de Le{\'o}n, I{\~n}igo Apaolaza, Alfonso Valencia, and Francisco~J
  Planes.
\newblock On the inconsistent treatment of gene-protein-reaction rules in
  context-specific metabolic models.
\newblock {\em Bioinformatics}, 36(6):1986, 2020.

\bibitem{velivckovic2017graph}
Petar Veli{\v{c}}kovi{\'c}, Guillem Cucurull, Arantxa Casanova, Adriana Romero,
  Pietro Lio, and Yoshua Bengio.
\newblock Graph attention networks.
\newblock {\em arXiv preprint arXiv:1710.10903}, 2017.

\bibitem{galuzzi2022best}
Bruno~G Galuzzi, Luca Milazzo, and Chiara Damiani.
\newblock Best practices in flux sampling of constrained-based models.
\newblock In {\em International Conference on Machine Learning, Optimization,
  and Data Science}, pages 234--248. Springer, 2022.

\bibitem{xiong2021optimized}
Kai Xiong, Karen~Julie la~Cour~Karottki, Hooman Hefzi, Songyuan Li, Lise~Marie
  Grav, Shangzhong Li, Philipp Spahn, Jae~Seong Lee, Ildze Ventina, Gyun~Min
  Lee, et~al.
\newblock An optimized genome-wide, virus-free crispr screen for mammalian
  cells.
\newblock {\em Cell reports methods}, 1(4), 2021.

\bibitem{loshchilov2017decoupled}
Ilya Loshchilov and Frank Hutter.
\newblock Decoupled weight decay regularization.
\newblock {\em arXiv preprint arXiv:1711.05101}, 2017.

\bibitem{grover2016node2vec}
Aditya Grover and Jure Leskovec.
\newblock node2vec: Scalable feature learning for networks.
\newblock In {\em Proceedings of the 22nd ACM SIGKDD international conference
  on Knowledge discovery and data mining}, pages 855--864, 2016.

\end{thebibliography}

\appendix

\section{Mass Flow Graph}
\begin{figure}[h!]
    \centering
    \includegraphics[width=0.4\textwidth]{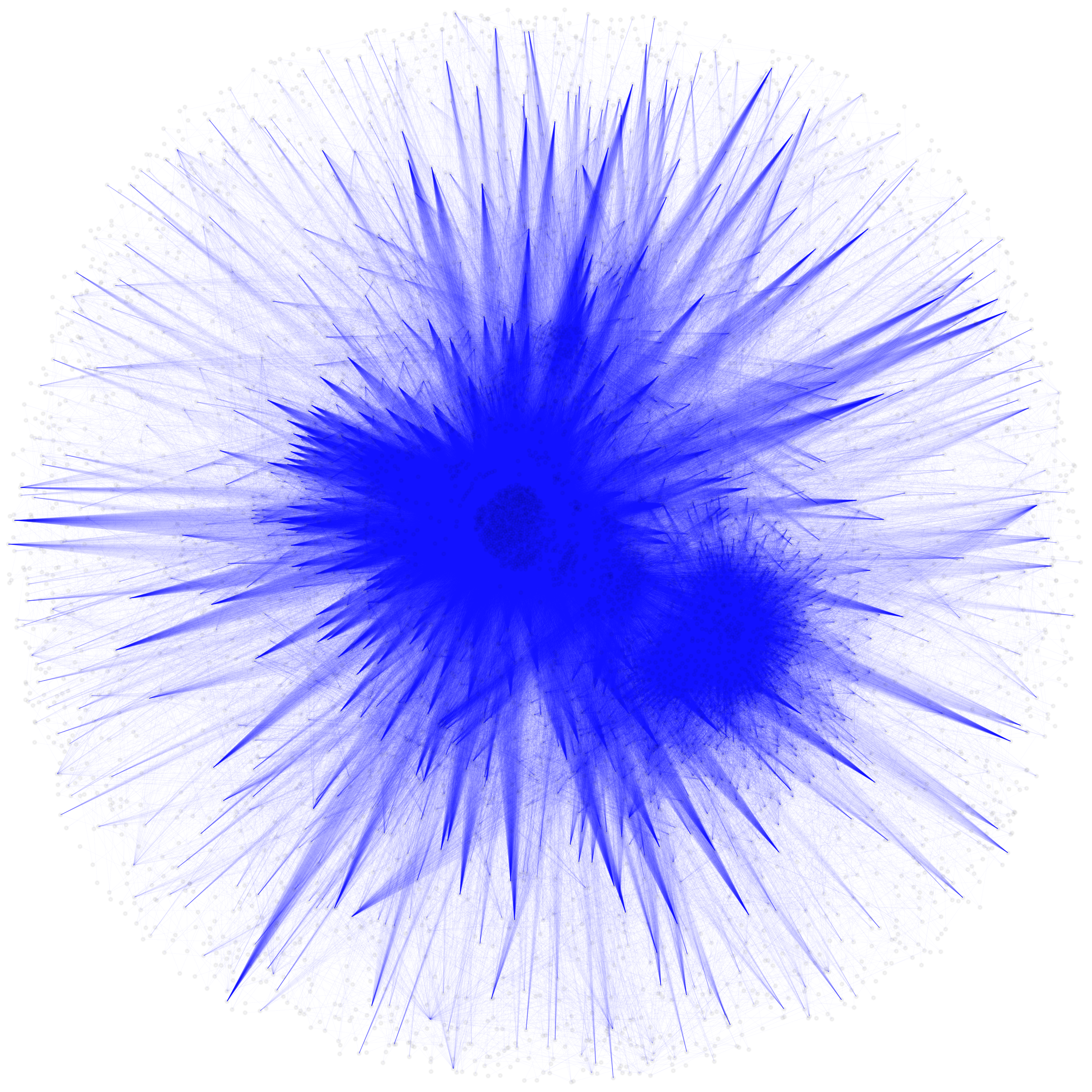}
    \caption{Mass flow graph (MFG) representation of the CHO cell metabolic reaction network, constructed using flux sampling data, containing 4,733 nodes (reactions) and 335,024 edges (interactions).}
    \label{fig:MFG}
\end{figure}

\section{Hyperparameter Optimisation}
Optimal hyperparameter selection for FluxGAT was crucial to enhance its learning efficiency and predictive accuracy. Our objective was to improve the model's generalisation capability and to ensure a balance between computational efficiency and performance. We employed a grid search strategy to identify the optimal hyperparameters for FluxGAT that maximised the effectiveness of the GAT architecture on the task of biological essentiality prediction. The numerical ranges were carefully chosen based on a combination of empirical testing and domain knowledge.

\begin{table}[h!]
\centering
\begin{tabular}{@{}rccc@{}}
\toprule
                              & Hyperparameter                   & Search Space  & Final Value          \\ \midrule
\multirow{5}{*}{\rotatebox{90}{Architecture}} & Number of message passing layers & {[}1-5{]}     & 2                    \\
                              & Number of attention heads        & {[}1-3{]}     & 2                    \\
                              & Dimension of embedding layer     & {[}25-300{]}  & 300                  \\
                              & Dimension of hidden layers       & {[}10-150{]}  & 150                  \\
                              & Activation function threshold    & {[}0.5-0.7{]} & 0.65                 \\ \midrule
\multirow{5}{*}{\rotatebox{90}{Training}}     & Base learning rate               & -             & 0.001                \\
                              & Learning rate decay              & -             & 0.01                 \\
                              & Optimisation algorithm           & -             & AdamW                \\
                              & Max epochs                 & -             & 200                  \\
                              & Regression loss                  & -             & Binary Cross Entropy \\ \bottomrule
\end{tabular}
\vspace{2mm}
\caption{\textbf{Overview of FluxGAT Hyperparameters and Training.} Summary of the search space and finalised values for the hyperparameters of the FluxGAT architecture, along with key training parameters.}
\label{tab:hyperparameters}
\end{table}

\section{Node Embedding Space}
\begin{figure}[h!]
    \centering
    \begin{subfigure}{0.45\columnwidth}
        \includegraphics[width=0.9\columnwidth]{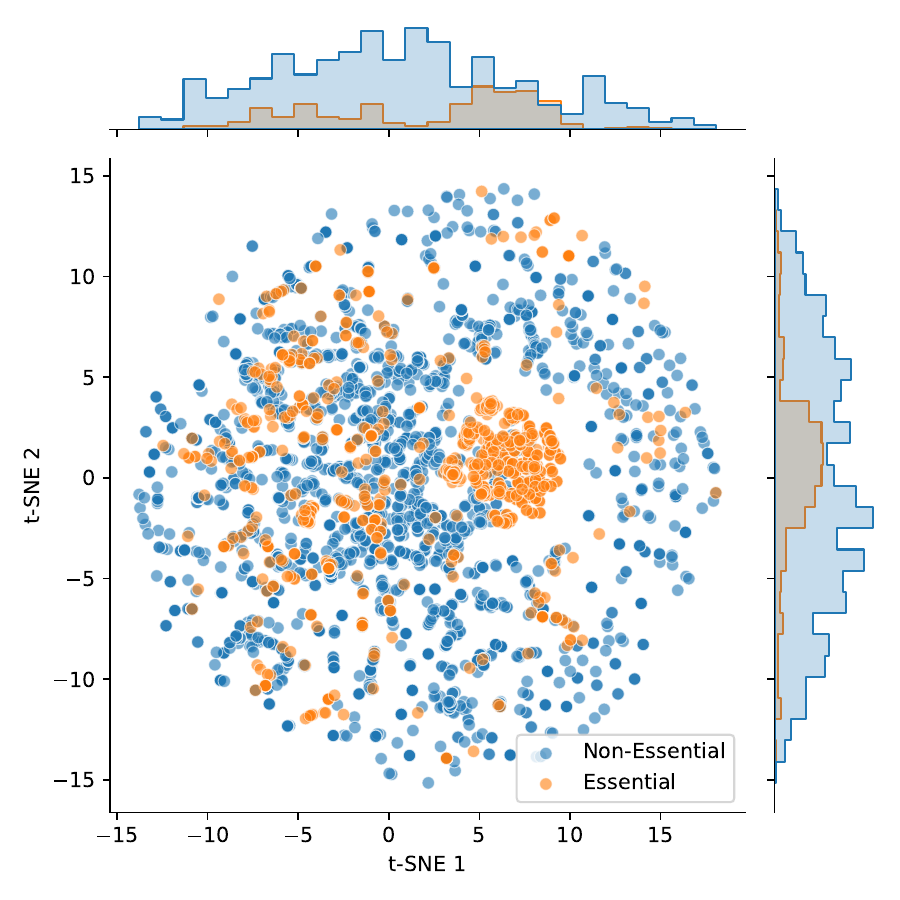}
        \caption{}
        \label{fig:input}
    \end{subfigure}
    \hspace{0.5cm} 
    \begin{subfigure}{0.45\columnwidth}
        \includegraphics[width=0.9\columnwidth]{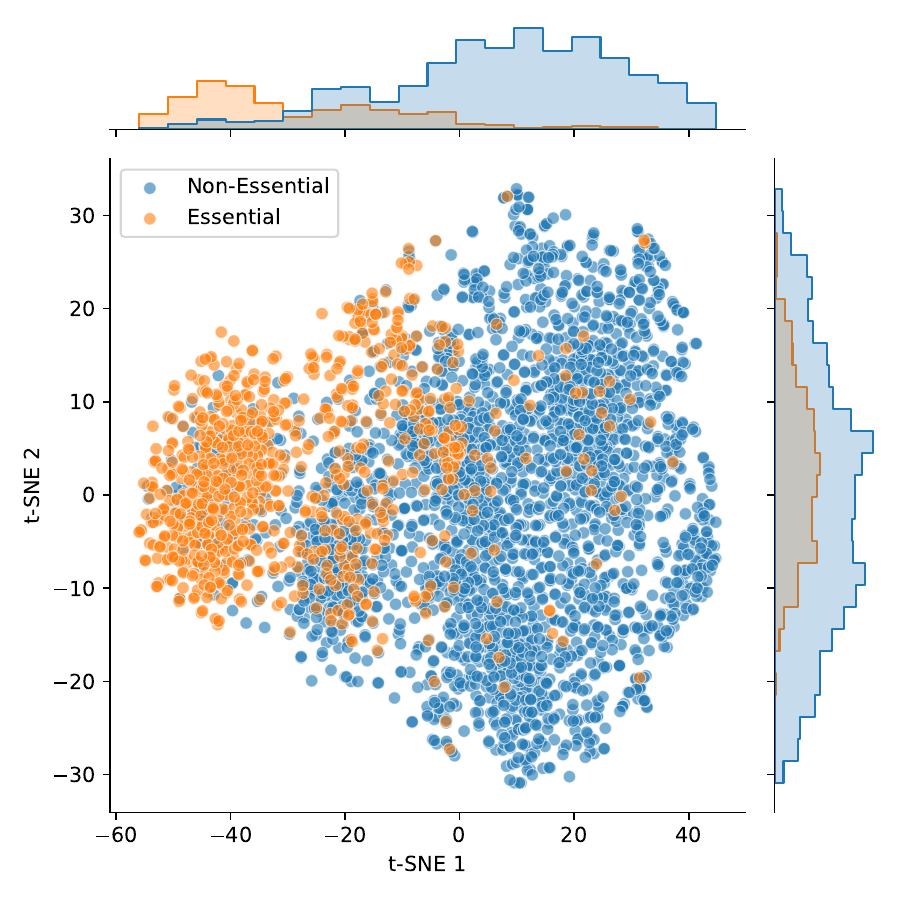}
        \caption{}
        \label{fig:output}
    \end{subfigure}
    \caption{t-SNE visualisations contrasting the structure of input features and learned node embeddings. (a) t-SNE reduction of the initial input features for each reaction, colour-coded as \textit{essential} and \textit{non-essential}, with marginal histograms depicting the distribution of t-SNE components for each class. (b) t-SNE reduced high-dimensional embeddings generated by FluxGAT, highlighting the model's ability to refine and separate nodes based on their essentiality through learned representations. Both plots used a perplexity of 30.}
    \label{fig:embeddings}
\end{figure}

For an exploration of the feature transformations enabled by FluxGAT, Figure \ref{fig:embeddings} illustrates the evolution from initial input features to refined embeddings, highlighting the model's effectiveness in distinguishing between essential and non-essential nodes. Future work will focus on understanding the underlying mechanisms and features that drive this transformation, aiming to identify the specific attributes and interactions FluxGAT leverages to classify essentiality.

\section{Computational Resources}
Simulations were conducted using a high-performance computing (HPC) cluster, which is equipped with the following node types:

\begin{itemize}
    \item Lenovo nx360 m5 compute nodes, each featuring two Intel E5-2680 v4 (Broadwell) CPUs with 14 cores at 2.4 GHz, and 128 GiB of RAM.

    \item High-memory nodes, each with 512 GiB of RAM.

    \item GPU-equipped nodes, each housing two graphics processing units.
\end{itemize}

For the experiments conducted, every node type was strategically utilised, based on the specific requirements of the job and the experiment phase in question. High-memory nodes were specifically employed for flux sampling of the iCHO2291 GSMM, whereas GPU nodes facilitated both the construction of MFGs and the training/evaluation of FluxGAT along with other GNN models. All remaining tasks were efficiently executed on the standard compute nodes.

\section{Code Availability}
To further support the reproducibility and transparency of our work, the complete source code for FluxGAT, including Python code and detailed instructions necessary for replication, is available on GitHub: \url{https://github.com/kierensharma/FluxGAT}. This version of the code has been modified to allow execution on personal computers, although users should note that this may significantly increase processing time compared to running the code on an HPC cluster.

\end{document}